# Modulation of the Nernst Thermoelectrics by Regulating the Anomalous Hall and Nernst Angles

Meng Lyu#, Junyan Liu#, Jianlei Shen, Shen Zhang, Yang Liu, Jinying Yang, Yibo Wang, Yiting Feng, Binbin Wang, Hongxiang Wei, Enke Liu*

Dr. M. Lyu, Dr. J. Liu, Dr. S. Zhang, Y. Liu, J. Yang, Y. Wang, Y. Feng, B. Wang, Prof. H. Wei, Prof. E. Liu
Beijing National Laboratory for Condensed Matter Physics,
Institute of Physics, Chinese Academy of Sciences, Beijing 100190, China
E-mail: ekliu@iphy.ac.cn

Dr. J. Shen
Key Laboratory of Magnetic Molecules and Magnetic Information Materials of Ministry of Education & Research Institute of Materials Science, Shanxi Normal University, Taiyuan 030000, China

**Abstract**

**The large anomalous Nernst effect in magnetic Weyl semimetals is one of the most intriguing transport phenomena, which draws significant attention for its potential applications in topological thermoelectrics. Despite frequent reports of substantial anomalous Nernst conductivity (ANC), methods to optimize Nernst thermoelectrics remain limited. Our research reveals that the magnitude of the ANC is directly related to the sum of the anomalous Nernst and Hall angles. While the sign of the anomalous Hall angle is relatively stable in a certain material, the sign of the anomalous Nernst angle can be intrinsically tuned. Therefore, the ANC can be effectively optimized by regulating these angles to work in concert. This finding is verified by experimental modulation from iron-doped magnetic topological material $Co_3Sn_2S_2$. Additionally, we observed a robust $T\ln T$ scaling law of the ANC over the temperature range of 40 to 140 K in all studied samples, suggesting an intrinsic origin of the ANC. Considering the common opposite sign of the anomalous Nernst and Hall angles in many magnetic topological materials, our research offers an applicable scheme for optimizing the Nernst thermoelectrics.**

## 1. Introduction

In recent years, as topological quantum materials widely come into people's field of vision, the Berry phase based anomalous phenomena have attracted increasing interest. Of particular interest are the anomalous Hall effect (AHE) and the anomalous Nernst effect (ANE).[1-5] Most reported anomalous Nernst conductivity (ANC, $\alpha_{yx}^{A}$) in the magnetic topological materials are one or two orders of magnitude larger than the conventional ferromagnetic materials (~ 0.001 - 0.1 A $m^{-1}$ $K^{-1}$) because of the large intrinsic Berry curvature, leading to a great potential in the zero-field transverse thermoelectric applications.[6, 7] Currently, more and more magnetic topological materials with large ANC are being discovered, such as $Co_3Sn_2S_2$ ($|\alpha_{yx}^{A}|_{max}$ ~ 3-4 A $m^{-1}$ $K^{-1}$),[5, 8, 9] $Co_2MnGa$ ($|\alpha_{yx}^{A}|_{max}$ ~ 4 A $m^{-1}$ $K^{-1}$),[10, 11] $UCo_{0.8}Ru_{0.2}Al$ ($|\alpha_{yx}^{A}|_{max}$ ~ 15 A $m^{-1}$ $K^{-1}$),[12] $Fe_3Ga(Al)$ ($|\alpha_{yx}^{A}|_{max}$ ~ 3.5-5.2 A $m^{-1}$ $K^{-1}$),[13] $Fe_3Sn$ ($|\alpha_{yx}^{A}|_{max}$ ~ 2.4 A $m^{-1}$ $K^{-1}$),[14] $YbMnBi_2$ ($|\alpha_{yx}^{A}|_{max}$ ~ 10 A $m^{-1}$ $K^{-1}$),[15] and MnBi ($|\alpha_{yx}^{A}|_{max}$ ~ 44 A $m^{-1}$ $K^{-1}$).[16] Despite the discovery of numerous materials with large ANC, there has been limited experimental research on how to improve their thermoelectric performance. Ding *et al* reported an increase in the intrinsic ANC resulting from disorder in their experimental approach.[8] Meanwhile, Minami *et al.* and Ivanov *et al.* separately indicated through theoretical calculations that the ANC can be improved by stationary points in nodal lines and coactive-staggered feature in anomalous Hall conductivity.[17, 18]

The magnetic topological semimetals are characterized by Weyl nodes or nodal lines, which generates large Berry curvature ($\Omega$). From Xiao's work,[1] the relation between $\Omega$ and the intrinsic $\alpha_{yx}^{A}$ are expressed as $\alpha_{yx}^{A} = -\frac{e}{T\hbar}\int\frac{dk^3}{2\pi^3}\Omega_{n,z}(k)s_{nk}$, where $s_{nk} = (\epsilon_{n,k}-\mu)f_{n,k} + k_B T\ln[1+e^{-(\epsilon_{n,k}-\mu)/k_B T}]$, e, $\hbar$, $k_B$, and $\mu$ are the elementary charge, the reduced Planck constant, the Boltzmann constant, and the chemical potential respectively, $s_n$ represents the weight factor of Berry curvature for the ANC. It is evident that the ANC is closely linked to the Berry curvature, which explains the fact that the ANC is remarkable in magnetic topological materials. Besides, the chemical potential in the weight function $s_n$ also plays an essential role in determining the value of ANC, so tuning the chemical potential could be a good way to improve ANC. However, the mechanisms by which chemical potential tuning facilitates the enhancement of ANC remain elusive. On the other hand, due to its close connection with Berry curvature, probing ANC is an efficient technique for revealing the topological band structure and the Berry curvature in Weyl semimetals.[19-21]

$Co_3Sn_2S_2$ has been recognized as the first magnetic Weyl semimetal, which belongs to the Shandite family of compounds with a rhombohedral structure (space group R-3m), in which the quasi-2D $Co_3Sn$ layers are sandwiched between S atoms along the *c*-axis (Figure 1a). The enlarged $Co_3Sn$ layer in Figure 1a clearly shows that the Co atoms form a kagome lattice, where the small amount of Fe dopants substitute for Co atoms. Theoretical calculations and ARPES measurements reveal the presence of Weyl points and nodal lines close to the Fermi level,[4, 22] suggesting the very large Berry curvature in this compound. STM and optical spectroscopy results show that it is a correlated kagome magnet with flat bands around the Fermi level. [23, 24] The large Berry curvature, electronic correlations and flat bands are considered to be benefit to the improvement of the ANC.[12] In fact, previous work has already reported that $Co_3Sn_2S_2$ exhibits a substantial ANC. In addition, some other studies have shown that the extrinsic contribution could play an important role in the anomalous Hall conductivity (AHC) of the magnetic topological semimetals, [25, 26] but there is currently rare research on how extrinsic contribution affects the ANC. Therefore, the chemical doping of $Co_3Sn_2S_2$ could be an optimal platform to investigate the optimization method of the ANC whilst identifying the contribution of extrinsic scattering effects to the ANC.

In the present study, we have carried out a comprehensive analysis of the transverse thermoelectric conductivity for the magnetic Weyl semimetal $Co_{3-x}Fe_xSn_2S_2$ with $x = 0$, 0.05, 0.10, 0.15 and 0.20 by measuring the longitudinal electric resistivity, Hall effect, Seebeck coefficient, and Nernst effect. The thermoelectric measurements schematic configuration is shown in Figure 1b. Firstly, we present that the maximum ANC of $Co_{3-x}Fe_xSn_2S_2$ can be enhanced by 82.4% from 1.59 to 2.9 A $m^{-1}$ $K^{-1}$ when the Fe content is increased to 0.15. Subsequently, we reveal that the ANC of these compounds is derived from the intrinsic Berry curvature contribution, as evidenced by the observation of typical $T\ln T$ behavior and the high degree of agreement with theoretical calculations. Finally, through a comprehensive analysis of the constituent elements of the ANC, we demonstrate that the anomalous Nernst angle can be inherently modified to align with the anomalous Hall angle, thereby significantly enhancing the ANC.

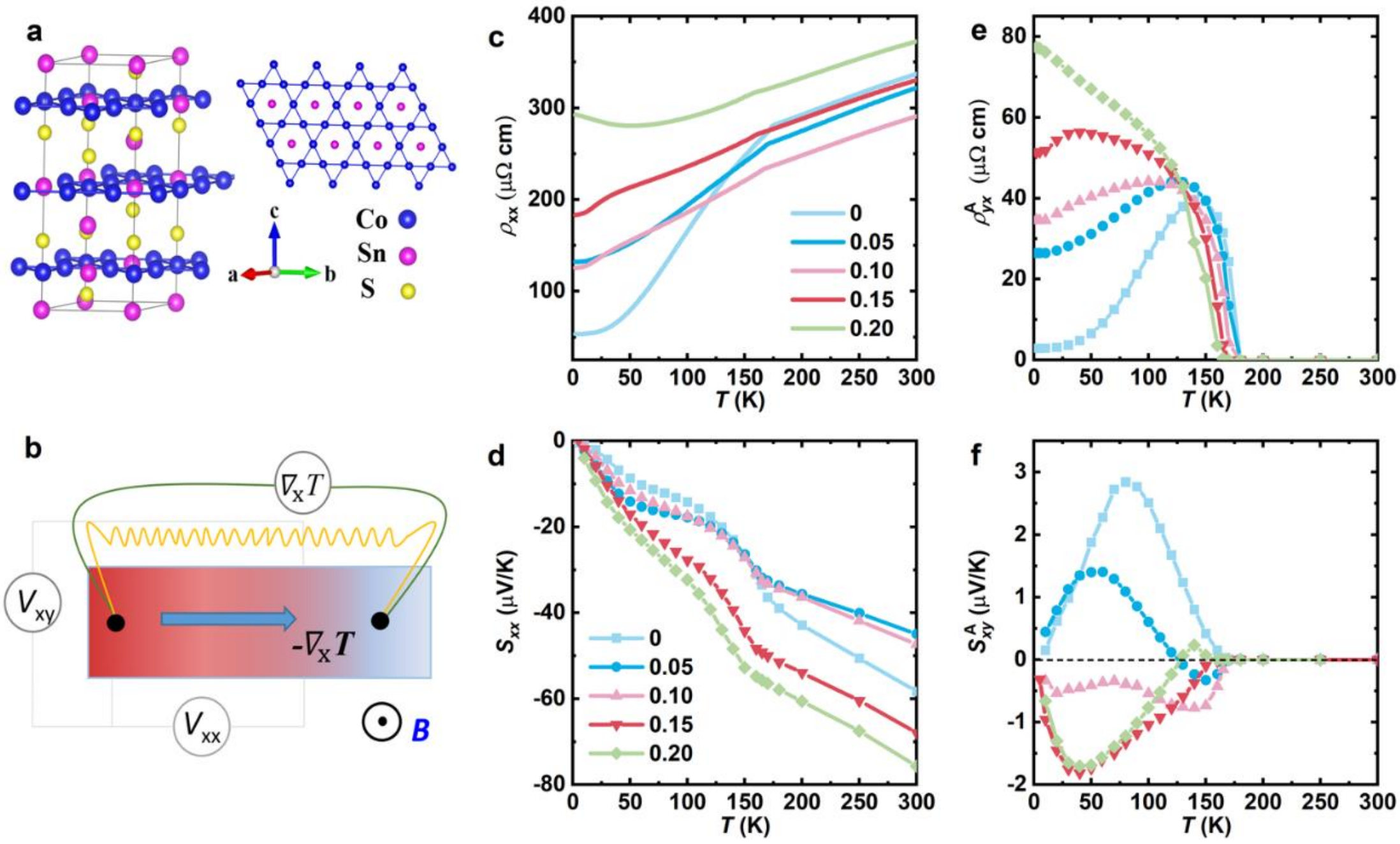

**Figure 1.** Crystal structure and transport properties of $Co_{3-x}Fe_xSn_2S_2$. a) Crystal structure of $Co_3Sn_2S_2$ with quasi-2D $Co_3Sn$ layers, where the Co atoms form a kagome lattice and the doped Fe atoms replace some of the Co atoms. b) Schematic configuration of thermoelectric measurements, where the heat current from a resistance chip heater generates the temperature gradient $\nabla_x T$, the longitudinal Seebeck voltage $V_{xx}$ and the transverse Nernst voltage $V_{yx}$ in the magnetic field. c) Temperature dependent electrical resistivity $\rho_{xx}(T)$ with $I$ // $a$. d) Temperature dependent Seebeck coefficient $S_{xx}$ $(T)$. e) Large Hall resistivity estimated from $\rho_{yx}(B)$ curves at different temperatures. f) Summarized $S_{yx}^{A}$ as a function of temperature.

## 2. Results and Discussion

Figure 1c displays the temperature dependent longitudinal resistivity $\rho_{xx}$ at zero field with $I$ // $a$. $\rho_{xx}$ ($T$) exhibits a clear kink at the Curie temperature $T_C$, which can be accurately estimated from the derivative $\rho_{xx}$ ($T$) curves (see details in Figure S2, Supplementary Information). The estimated $T_C$ here for different Fe-content samples are in agreement with previous work.[25] The kink features of the ferromagnetic ordering are also observed in the Seebeck coefficient $S_{xx}$ at the same temperature, see in Figure 1d. The negative $S_{xx}$ in the whole temperature range indicates that the electrons are the dominant carriers for $Co_{3-x}Fe_xSn_2S_2$. Furthermore, we notice that $S_{xx}$ is significantly enhanced around the room temperature with increasing Fe doping, suggesting that $Co_{3-x}Fe_xSn_2S_2$ could be a good longitudinal thermoelectric material with more Fe doping at higher temperatures. Figure 1e displays the temperature dependence of the anomalous Hall resistivity $\rho_{yx}^{\mathrm{A}}$(estimated from the detailed $\rho_{yx}$ vs $H$ plots, see in Figure S3), which exhibits a giant value compared to other ferromagnetic materials, revealing the giant anomalous Hall conductivity and anomalous Hall angle in $Co_{3-x}Fe_xSn_2S_2$.

After a brief introduction to $\rho_{xx}$ ($T$), $S_{xx}$ ($T$), and $\rho_{yx}^{\mathrm{A}}(T)$ of $Co_{3-x}Fe_xSn_2S_2$, we turn to the anomalous Nernst effect $S_{yx}^{\mathrm{A}}(T)$ at zero magnetic field. Considering the maximum coercive fields of these compounds close to 0.3 T, we show only the data between -1 T and +1 T to estimate $S_{yx}^{\mathrm{A}}(T)$. As seen from the curves of Figure S4, $S_{\mathrm{yx}}$ ($H$) undergoes a sign change with increasing Fe content, which is quite different from the Hall effect and the reported magnetization curves.[25, 27] In more details, $S_{yx}$ changes sign at $x$ = 0.05 and 0.20 with increasing temperature, marked by the pink arrow in Figures S4b and S4e. This temperature-dependent sign change behavior of the Nernst signal was also observed in the pristine $Co_3Sn_2S_2$ nanoribbons,[28] which was attributed to the magnetic fluctuations around $T_C$. By extrapolating $S_{yx}$ ($H$) to zero magnetic field, we obtained $S_{yx}^{\mathrm{A}}(T)$ of the variable Fe content shown in Figure 1f. This plot shows clearly the sign change behavior of $S_{yx}^{\mathrm{A}}$ with Fe doping. The sign change behavior of $S_{yx}^{\mathrm{A}}$ by doping is rare so far.[29, 30] In Yang's report of $Fe_{3-\delta}GeTe_2$,[29] $S_{yx}^{\mathrm{A}}$ also undergoes a sign change with decreasing Fe vacancy, and they proposed that it is attributed to the chemical potential shift in the electronic bands. It should be noted that Liu *et al* have also recently measured the Nernst effect for the Fe-doped $Co_3Sn_2S_2$,[31] but their experimental results are quite different from ours. As they mentioned, the discrepancy may be due to the different grown-flux method.

It is well known that the topological band structure can be understood by analyzing the temperature dependent ANC ($\alpha_{yx}^{\mathrm{A}}$), which is directly related to the measured $\rho_{xx}$, $\rho_{yx}^{\mathrm{A}}$, $S_{xx}$, and $S_{yx}^{\mathrm{A}}$,

$$\alpha_{yx}^{\mathrm{A}} = (S_{yx}^{\mathrm{A}}\rho_{xx} - S_{xx}\rho_{yx}^{\mathrm{A}})/({\rho_{xx}}^2 + {\rho_{yx}^{\mathrm{A}}}^2) \tag{1}$$

Due to the longitudinal electrical conductivity $\sigma_{xx} = \rho_{xx}/({\rho_{xx}}^2 + {\rho_{yx}^{\mathrm{A}}}^2)$ and the anomalous Hall conductivity $\sigma_{yx}^{\mathrm{A}} = -\rho_{yx}^{\mathrm{A}}/({\rho_{xx}}^2 + {\rho_{yx}^{\mathrm{A}}}^2)$, the ANC could also be expressed as

$$\alpha_{yx}^{\mathrm{A}} = S_{xx}\sigma_{yx}^{\mathrm{A}} + S_{yx}^{\mathrm{A}}\sigma_{xx} \tag{2}$$

Based on the measured values of $\rho_{xx}$, $\rho_{yx}^{\mathrm{A}}$, $S_{xx}$, and $S_{yx}^{\mathrm{A}}$, the ANC can be easily obtained using Eq. (1). Figure 2a presents the temperature dependence of the ANC for $Co_{3-x}Fe_xSn_2S_2$. The plot shows that the measured result of the pristine $Co_3Sn_2S_2$ is consistent with the previously reported work using a single crystal also grown from the high-temperature flux method.[8] Furthermore, it is observed that the ANC stays a large value over a wider temperature range with Fe doping and the maximum value is significantly amplified by 82.4%, from 1.63 A m$^{-1}$ K$^{-1}$ at $x = 0$, increasing to 2.9 A m$^{-1}$ K$^{-1}$ at $x = 0.15$, and then decreasing at $x = 0.20$ , see the inset plot of Figure 2a. Such a large enhancement of the ANC has great potential for optimizing the transverse thermoelectric merit. Figures 2b and 2c display the two components of the ANC describing by the middle two terms of Eq. (2). It is clear that the first term remains positive for all samples. However, the second term changes from negative to positive as more Fe is doped, resulting in an increase in the total ANC.

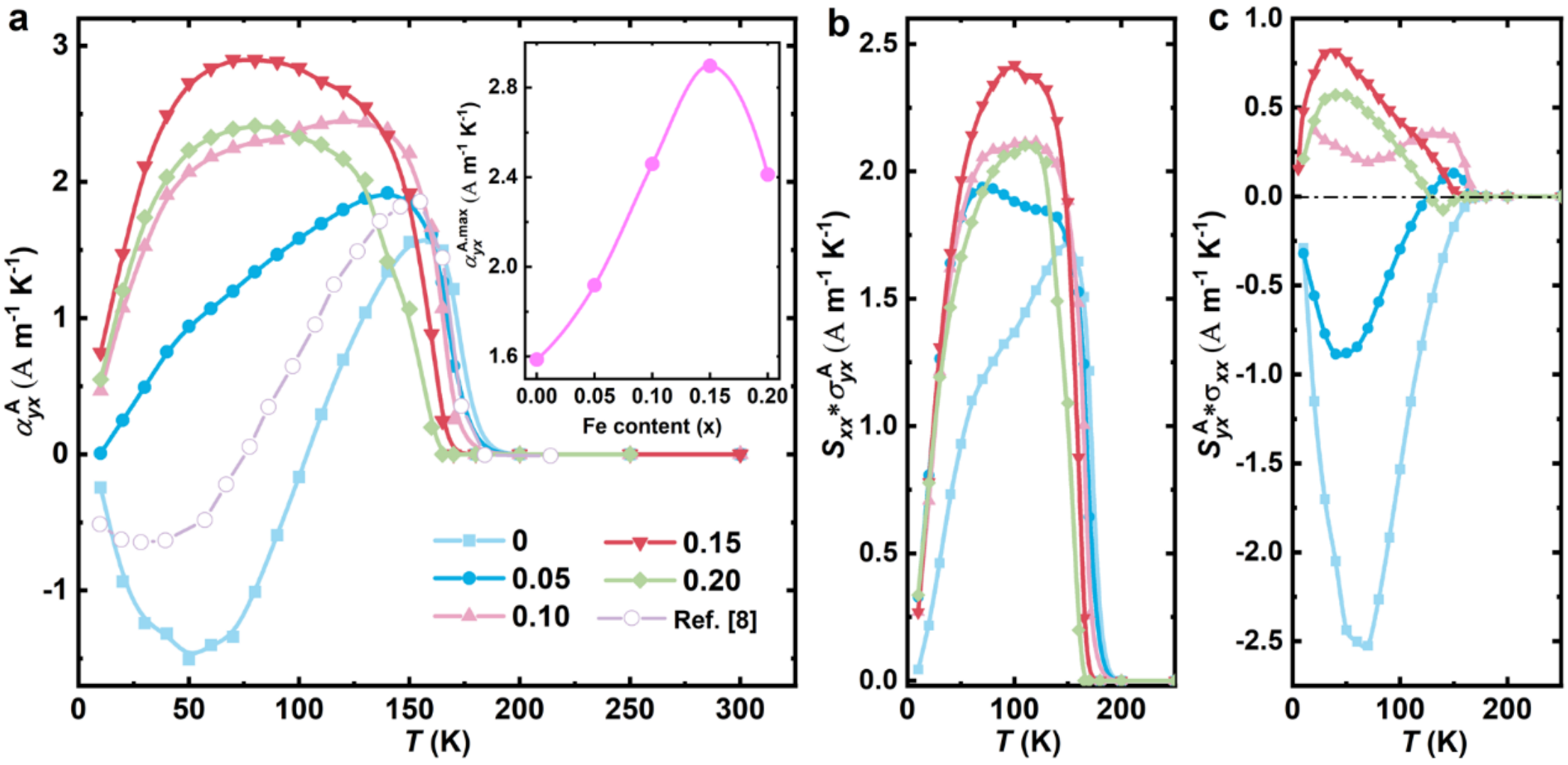

**Figure 2.** Anomalous Nernst conductivity of $Co_{3-x}Fe_xSn_2S_2$. a) The anomalous Nernst conductivity $\alpha_{yx}^{\mathrm{A}}$ was calculated from the measured $\rho_{xx}$, $\rho_{yx}^{\mathrm{A}}$, $S_{xx}$, and $S_{yx}^{\mathrm{A}}$. It also shows the data of pristine $Co_3Sn_2S_2$ from Ref. [8], which is consistent with our crystal. The inset illustrates the maximum ANC at varying Fe contents. One can immediately see that $\alpha_{yx}^{\mathrm{A}}$ is largely enhanced as the Fe content increases to 0.15. b), c) The temperature dependence of two ANC components, $S_{xx}\sigma_{yx}^{\mathrm{A}}$ and $S_{yx}^{\mathrm{A}}\sigma_{xx}$. The total $\alpha_{yx}^{\mathrm{A}}$ is the sum of two parts.

The previous research[25] has effectively distinguished between the intrinsic and extrinsic contributions to the anomalous Hall conductivity utilizing the Tian-Ye-Jin (TYJ) model[32, 33] in $Co_{3-x}Fe_xSn_2S_2$, and found that the extrinsic contribution markedly improves the AHC when subjected to slight Fe doping. Here, we have conducted the same separation on the AHC of the studied samples in Figure 3a. In line with the previously reported results,[25] it can be observed that the extrinsic contribution gradually increases and becomes dominant in the total value of AHC when $x \geq 0.1$. Meanwhile, as mentioned above, the maximum ANC is also largely enhanced with a small Fe doping, as shown on the right axis of Figure 3a. Therefore, it would be interesting to find out how chemical doping tunes the ANC and whether the extrinsic scattering mechanism also contributes significantly to the ANC in Fe-doped $Co_3Sn_2S_2$.

Figure 3b shows the temperature dependence of $\alpha_{yx}^{\mathrm{A}}/T$ for different magnetic topological materials. As seen in Figures 3b and 3c, it can be clearly observed that all $\alpha_{yx}^{\mathrm{A}}/T$ curves of $Co_{3-x}Fe_xSn_2S_2$ follow the relationship of -ln$T$ in the temperature range from 40 to 140 K. This is the same behavior that has been observed in many other magnetic materials, such as $Co_2MnGa$, $Fe_3Ga$, and $YbMnBi_2$, CoMnSb.[10, 12-15, 34] As shown in Figure 3b, the -ln$T$ feature is marked by the pink dotted lines for the reported magnetic topological materials, suggesting that the large ANC of $Co_{3-x}Fe_xSn_2S_2$ could also originate from the intrinsic topological bands. The scaling logarithmic temperature behavior of $\alpha_{yx}^{\mathrm{A}}/T$ well below the Curie temperature appears to be a robust law for most of the magnetic topological materials, implying a common origin of Berry curvature. In the Ref. [17], Susumu *et al.* proposed a good explanation for this peculiar temperature dependence of $\alpha_{yx}^{\mathrm{A}}/T$. They claimed that when the chemical potential is in the proximity of the van Hove singularities within the nodal line, then $\alpha_{yx}^{\mathrm{A}}/T$ increases substantially. As an example, the ANC of $Co_3Sn_2S_2$ was calculated based on the density functional theory, as shown by the filled dotted lines in Figure 3c. The behavior of $\alpha_{yx}^{\mathrm{A}}/T$ in their calculations is linear with ln$T$, and transitions from negative to positive at low temperatures with decreasing chemical potential, which is strikingly consistent in magnitude and sign with our measured results (comparison in Figure 3c). The good agreement between their calculations and our experiments suggests that the large ANC of $Co_{3-x}Fe_xSn_2S_2$ is primarily a result of the intrinsic topological band, while the small amount of Fe-hole doping serves to lower the chemical potential energy.

To further clarify the intrinsic contribution of the large ANC in the studied compounds, we have also performed a density function theory calculation on the pristine $Co_3Sn_2S_2$. Theoretically, the AHC can be calculated using $\sigma_{yx}^{\mathrm{A}} = -\frac{\mathrm{e}}{T\hbar}\int\frac{\mathrm{dk}^3}{2\pi^3}\Omega_{\mathrm{n,z}}(k)f_{\mathrm{n,\,k}}$ at zero temperature, where $f_{\mathrm{n,k}}$ is the Fermi-Dirac distribution function. Combined with the Mott relation $\alpha_{yx}^{\mathrm{A}} = \frac{\pi^2 k_B^2 T}{3e}\frac{d\sigma_{yx}^{\mathrm{A}}}{d\mu}$, $\alpha_{yx}^{\mathrm{A}}/T$ can be obtained near the zero temperature. Figure 3d shows the energy dependence of the calculated $\alpha_{yx}^{\mathrm{A}}/T$ from the Mott relation. The change in $\alpha_{yx}^{\mathrm{A}}/T$ shifts from negative to positive as the chemical potential moves from a few hundred meV above the Fermi level to lower energies. This is consistent with the experimental results shown in Figures 3b and 3c, and Fe-hole doping is expected to shift the chemical potential downwards. In addition, the calculated ANC as a function of the chemical potential at $T$ = 50, 80 and 100 K is shown in Figure 3e. We find that the value of the measured ANC at the same temperature shown in Figure 3f is very close to the calculated one. Upon careful comparison with theoretical calculations, the experimental results obtained exhibit a high degree of consistency with the calculated values. This confirms that the ANC of $Co_{3-x}Fe_xSn_2S_2$ is mainly derived from the contribution of intrinsic topological bands, as opposed to the dominant contribution of the AHC, where the extrinsic contribution is significantly enhanced at higher Fe doping.

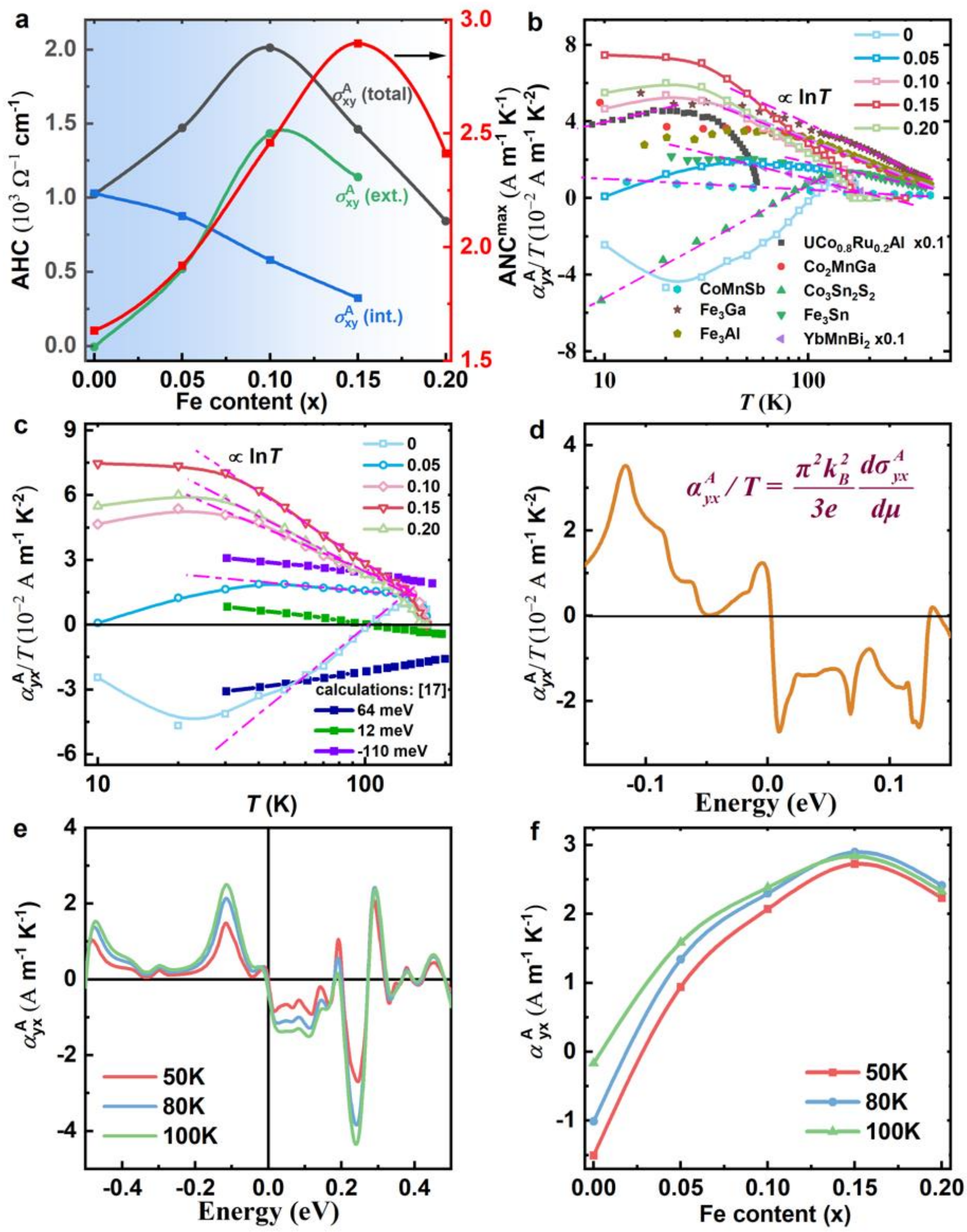

**Figure 3**. Theoretical and experimental comparisons of the ANC. a) The separation of intrinsic and extrinsic contributions to AHC by the TYJ model at $T = 2$ K. The right side shows the close trend of the maximum ANC to the total AHC. b) The observed $T\ln T$ behavior of $Co_{3-x}Fe_xSn_2S_2$. The point lines marked by pink dashed dots are the summary of the same $T\ln T$ scaling law for the reported magnetic topological materials. c) Susumu's calculation of $\alpha_{yx}^{\mathrm{A}}/T$ with chemical potential $\mu$ = 64, 12, and -110 meV. Both their calculated and our measured $\alpha_{yx}^{\mathrm{A}}/T$ show a logarithmic temperature behavior. d) The chemical potential dependent $\alpha_{yx}^{\mathrm{A}}/T$ calculated from the Mott relation. e) Calculated ANC as a function of energy for the pristine $Co_3Sn_2S_2$ at $T$ = 50, 80 and 100 K, whose behavior and magnitude are in good agreement with the experimental results at the same temperature in f).

In order to verify how chemical doping intrinsically tunes the ANC and to figure out a way to optimize the transverse thermoelectric merits, we conducted a more detailed analysis to the constitutive terms of the ANC. The ANC in Eq. (2) can be expressed as

$$\alpha_{yx}^{\mathrm{A}} = \alpha_{xx}\left(\tan\theta_H^{\mathrm{A}} + \tan\theta_N^{\mathrm{A}}\right) \quad (3)$$

Where $\alpha_{xx}$ denotes the longitudinal Peltier conductivity $\approx \sigma_{xx}S_{xx}$, $\tan\theta_H^{\mathrm{A}}$ is the anomalous Hall angle $\equiv \sigma_{yx}^{\mathrm{A}}/\sigma_{xx}$, and $\tan\theta_N^{\mathrm{A}}$ is the anomalous Nernst angle $\equiv S_{yx}^{\mathrm{A}}/S_{xx}$. For Eq. (3), one immediately recalls the small value of the ordinary Nernst effect in conventional metals due to the cancellation between the two angles $\theta_N$ and $\theta_H$, as called Sondheimer cancellation.[35, 36] Based on Eq. (3), the ANC is equal to multiplying the Peltier conductivity $\alpha_{xx}$ by the sum of $\tan\theta_H^{\mathrm{A}}$ and $\tan\theta_N^{\mathrm{A}}$. These two components of $Co_{3-x}Fe_xSn_2S_2$ are separately presented in Figures 4a and 4b. The inset of Figure 4b shows the variation of $(\tan\theta_H^{\mathrm{A}} + \tan\theta_N^{\mathrm{A}})$ with different Fe content at the selected temperature $T$ = 10, 50, 80 K. Although the absolute value of $(\tan\theta_H^{\mathrm{A}} + \tan\theta_N^{\mathrm{A}})$ increases to 0.78 at $x = 0.15$ and $T = 10$ K, the value of $\alpha_{\mathrm{yx}}^{\mathrm{A}}$ remains small due to the small size of $\alpha_{xx}$ at this temperature. In particular, we notice that the absolute value of $(\tan\theta_H^{\mathrm{A}} + \tan\theta_N^{\mathrm{A}})$ is small for the pristine $Co_3Sn_2S_2$ and gradually increases with increasing Fe doping, which can qualitatively explain the amplification behavior of the ANC. By further displaying the separated curves of $\tan\theta_H^{\mathrm{A}}$ and $\tan\theta_N^{\mathrm{A}}$ in Figure 4c and 4d, we find that these two terms exhibit opposite signs in the pristine $Co_3Sn_2S_2$ and thus cancel each other out, resulting in a relatively small value of $\alpha_{yx}^{\mathrm{A}}$. Nevertheless, because of the limited Fe doping, there occurs a sign change for $S_{yx}^{\mathrm{A}}$ as well as for $\tan\theta_N^{\mathrm{A}}$. The sign of $\tan\theta_N^{\mathrm{A}}$ changes from negative to positive and increases monotonically with increasing Fe doping, as shown in Figure 4d. In comparison, it can be observed that the sign of $\tan\theta_H^{\mathrm{A}}$ is consistently positive, as demonstrated in Figure 4c. This observation aligns with the assertion presented in our recent study that the sign of $\sigma_{yx}^{\mathrm{A}}$ is exclusively associated with the chirality of the Weyl fermions and exhibits heightened stability based on an effective Weyl model.[37] Since $\tan\theta_N^{\mathrm{A}}$ are tuned to bear the same sign with $\tan\theta_H^{\mathrm{A}}$, their sum gradually increases, leading to a large improvement in the ANC and the large values remain in a wide temperature range, see in Figure 2a. This result reveals that the ANC can be optimized by tuning $\tan\theta_H^{\mathrm{A}}$ and $\tan\theta_N^{\mathrm{A}}$ to the same sign to make them work together positively. Figure 4e illustrates the physical image of the sign change behavior of the intrinsic anomalous Nernst effect based on an effective single-Weyl model.[37] The effective single-Weyl model offers a simplified representation of the bands contributing to topological transport in a given magnetic topological material, wherein a single pair of Weyl nodes is used to describe the relevant physics. As confirmed above, the chemical potential is lowered with a slight Fe hole doping in $Co_3Sn_2S_2$, which causes the gap ($\Delta$) between the effective Weyl node and the Fermi energy to change the sign. This is followed by the Berry curvature also changing the sign, and finally the $S_{yx}^{\mathrm{A}}$. It should be noted that the ANC is directly related to the Berry curvature, while the anomalous Nernst signal is only indirectly related to it.

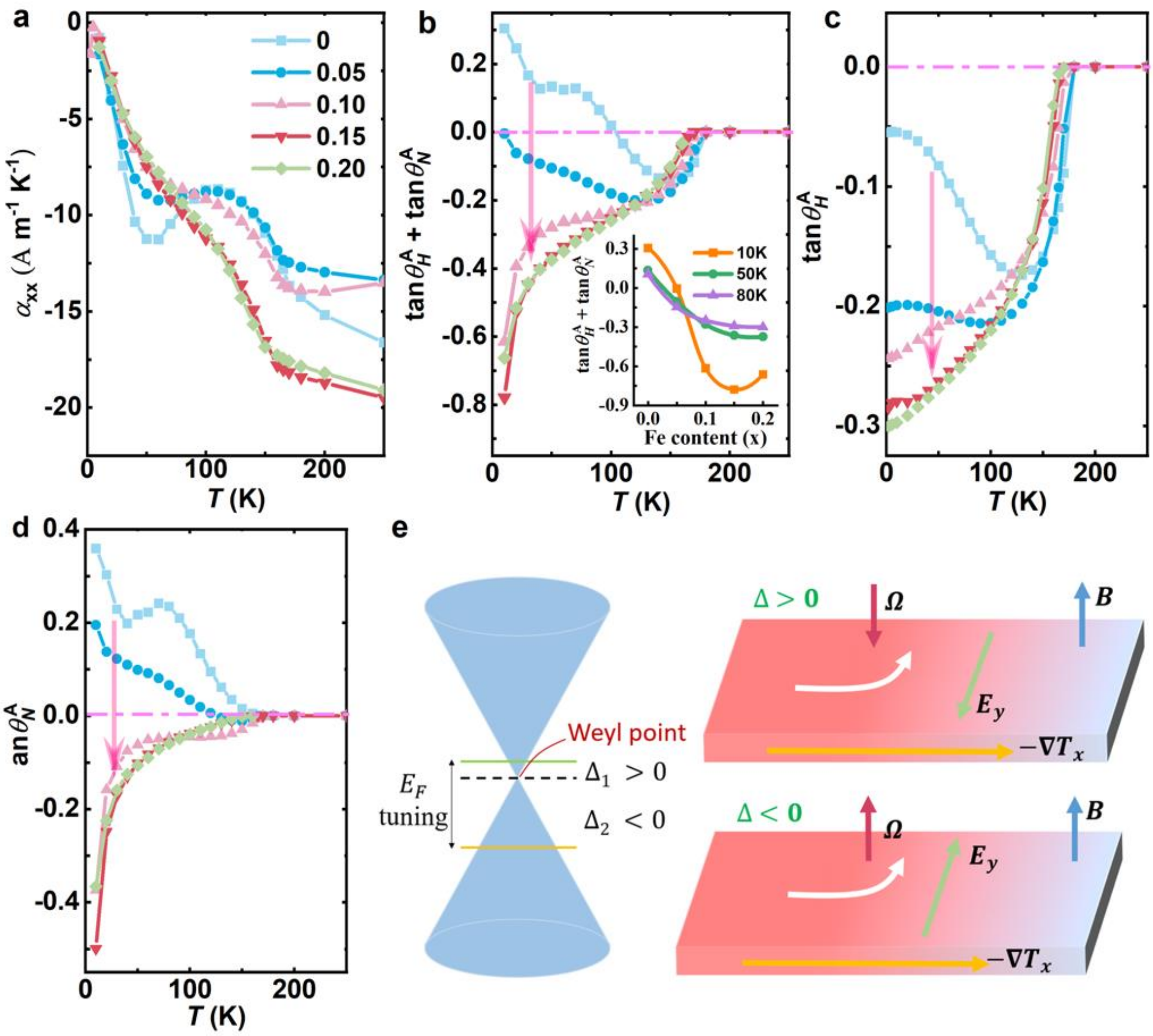

**Figure 4.** Detailed terms of the ANC in $Co_{3-x}Fe_xSn_2S_2$. a), b) The temperature dependent longitudinal Peltier conductivity $\alpha_{xx}$ and the sum of $\tan\theta_N^A$ and $\tan\theta_H^A$, the total ANC being the multiplication by these two terms. The inset shows the variation of $(\tan\theta_H^A + \tan\theta_N^A)$ with different Fe content at the selected temperature $T$ = 10, 50, 80 K. c), d) The temperature dependent $\tan\theta_H^A$ and $\tan\theta_N^A$. The sign of $\tan\theta_N^A$ changes from positive to negative with increasing Fe doping, while $\tan\theta_H^A$ always remains negative. e) Physical image of the sign change behavior of the intrinsic ANE with tuning of the chemical potential.

Actually, the cancellation behavior in the ANC caused by the opposite sign of $\tan\theta_H^A$ and $\tan\theta_N^A$ widely exists in many other magnetic topological materials. Figure 5 presents a summary of the maximum ANC for the reported magnetic topological systems in the upper panel. [10, 12-14, 16, 34, 38-42] The lower panel shows the corresponding $\tan\theta_H^A$ and $\tan\theta_N^A$, revealing a clear difference in sign between both in many systems. Specifically, with the largest ANC, MnBi, exhibits the same sign and a comparatively large value of $\tan\theta_H^A$ and $\tan\theta_N^A$. In contrast, $Fe_3GeTe_2$ has values of $\tan\theta_H^A$ and $\tan\theta_N^A$ that almost cancel each other out, resulting in a very small ANC. Based on the results presented in the green shadow of Figure 5 for $Co_{3-x}Fe_xSn_2S_2$, it is evident that the ANC is obviously improved by tuning $\tan\theta_H^A$ and $\tan\theta_N^A$ to work together positively, instead of cancelling each other out. Thus, it is possible to significantly enhance the ANC of systems with opposite signs of $\tan\theta_H^A$ and $\tan\theta_N^A$ by tuning them to the same sign through chemical doping or other ways. This provides a workable scheme to tune the ANC from the view of anomalous Nernst angle and anomalous Hall angle. Although the sign tuning picture of the intrinsic ANE is demonstrated above, it is important to stress that our tuning method of the ANC is broadly applicable, regardless of whether intrinsic or extrinsic contributions dominate the anomalous transverse transport.

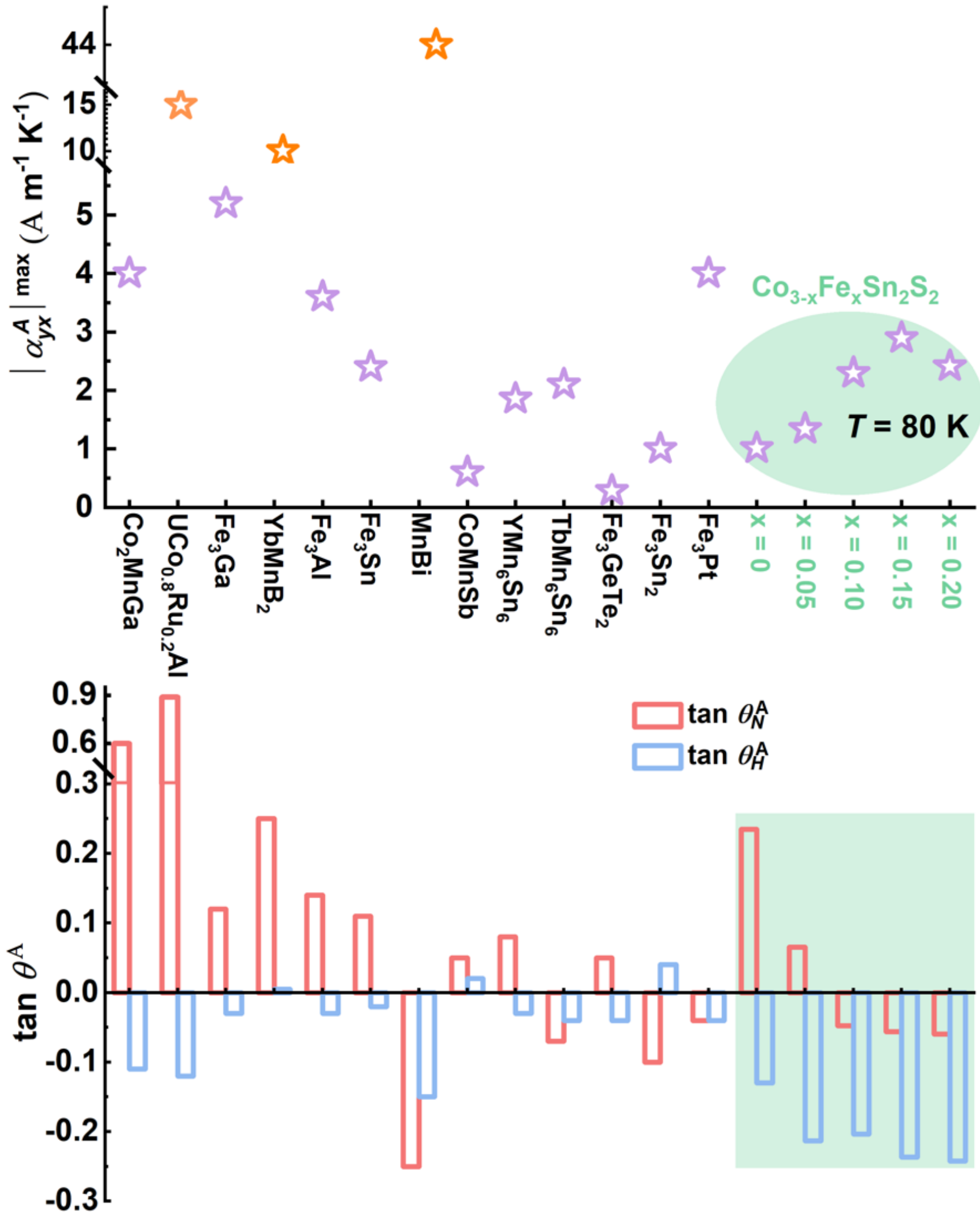

**Figure 5.** The upper panel shows the maximum ANC for the reported magnetic topological system. The lower panel displays the corresponding $\tan\theta_H^A$ and $\tan\theta_N^A$. The data in the green shadow indicates a significant amplification of the ANC (at $T$ = 80 K) for $Co_{3-x}Fe_xSn_2S_2$ when tuning $\tan\theta_H^A$ and $\tan\theta_N^A$ to the same sign. YbMnBi2 的 Bi 写错了。

## 3. Conclusion

In summary, we have systematically investigated the ANC in the iron-doped magnetic Weyl semimetal $Co_3Sn_2S_2$. We find that by positively modulating the anomalous Nernst angle and the anomalous Hall angle to the same sign, the maximum ANC can be increased by 82.4%. Through a detailed comparison with the theoretical calculations, we confirm that Fe-hole doping lowers the chemical potential, which leads to the sign change behavior of the anomalous Nernst angle and then intrinsically optimizes the ANC. Our research demonstrates that the intrinsic ANC in $Co_3Sn_2S_2$ is more robust than AHC when small chemical potential shifts are taken into account. Both the widely observed -$T\ln T$ dependent relationship and the robust intrinsic origin of the ANC suggest that the Nernst effect measurement could play an important role in uncovering the topological properties of magnetic topological materials. Furthermore, the tuning method of the ANC in this work could provide a valuable insight for material scientists to optimize the Nernst thermoelectrics of magnetic Weyl semimetals.

## 4. Experimental section

*Single Crystal Growth of $Co_{3-x}Fe_xSn_2S_2$*: Large single crystals of $Co_{3-x}Fe_xSn_2S_2$ were grown by high-temperature Pb and Sn mixed flux. The growing process and chemical composition analysis were described thoroughly in our recent work.[25]

*Electrical Transport Measurements*: The measured samples are cut and polished into a regular bar shape with dimensions of about 3×1×0.3 mm$^3$. The electrical transport measurements were carried out in the physical property measurement system (PPMS) between 2 K and room temperature. The standard six-probe method was use for the longitudinal resistivity and the Hall effect measurements with a current along the $a$-axis and magnetic fields parallel to the $c$-axis.

*Thermoelectric Transport Measurements*: The thermoelectric measurements including Seebeck coefficient and Nernst effect were employed with a steady-state method in the PPMS high vacuum environment. As shown in the schematic configuration of Figure 1b, the heat flow along the $x$-axis was supplied by a resistance chip heater, the longitudinal and transverse voltages $V_{xx}$ and $V_{yx}$ separately represent the Seebeck and Nernst signals, and the temperature gradient $\nabla_x T$ was measured by an A-B-A type thermocouple. Here $\nabla_x T = \Delta T/d$, $\Delta T$ is the temperature difference between the two thermocouple points attached to the sample, which was set to be 2%-3% of the sample temperature; $d$ is the distance of these two thermocouple points. The marked $x$-axis in the figure corresponds to the sample $a$-axis, and the actual image of this configuration is described in the supplementary information (Figure S1).

The Seebeck coefficient $S_{xx}$ and the Nernst signal $S_{yx}$ were estimated as $S_{xx} = V_{xx}/(L_x\nabla_x T)$ and $S_{yx} = V_{yx}/(L_y\nabla_x T)$, where $L_x$ and $L_y$ are the distance between the longitudinal and transverse voltage contacts, respectively. In order to eliminate the influence of the misalignment of the lead contacts, all the magnetoresistance (Seebeck signal) and Hall effect (Nernst effect) measurements were conducted by scanning both negative and positive magnetic fields. The same crystal was used for both the electrical and the thermoelectric transport measurements for the different Fe contents.

## Acknowledgements

This work was supported by the State Key Development Program for Basic Research of China (Nos. 2019YFA0704900, 2022YFA1403800, and 2022YFA1403400), the Fundamental Science Center of the National Natural Science Foundation of China (No. 52088101), the National Natural Science Foundation of China (Nos. 11974394, 12174426 and 12104280), the Strategic Priority Research Program (B) of the Chinese Academy of Sciences (CAS) (No. XDB33000000), the CAS Project for Young Scientists in Basic Research YSBR-057, the Synergetic Extreme Condition User Facility (SECUF), and the Scientific Instrument Developing Project of CAS (No. ZDKYYQ20210003).

## Conflict of Interest

The authors declare no conflict of interest.